# Electronics and Chemistry: Varying Single Molecule Junction Conductance Using Chemical Substituents

*Latha Venkataraman[1,4], Young S. Park[2,4], Adam C. Whalley[2,4] Colin Nuckolls[2,4],*

*Mark S. Hybertsen[3,4], Michael L. Steigerwald[2]*

[1]*Department of Physics,* [2]*Department of Chemistry,*
[3]*Department of Applied Physics and Applied Mathematics, and*

[4]*Center for Electron Transport in Molecular Nanostructures*

*Columbia University, New York, New York*

**Abstract**: We measure the low bias conductance of a series of substituted benzene diamine molecules while breaking a gold point contact in a solution of the molecules. Transport through these substituted benzenes is by means of non-resonant tunneling or super-exchange, with the molecular junction conductance depending on the alignment of the metal Fermi level to the closest molecular level. Electron donating substituents, which drive the occupied molecular orbitals up, increase the junction conductance while electron withdrawing substituents have the opposite effect. Thus for the measured series, conductance varies inversely with the calculated ionization potential of the molecules. These results reveal that the occupied states are closest to the gold Fermi energy, indicating that the tunneling transport through these molecules is analogous to hole tunneling through an insulating film.

Understanding the transport characteristics of molecules bonded between metal electrodes is of fundamental importance for molecular scale electronics.[1] It is well known that these transport characteristics are influenced by the intrinsic properties of the molecules, including their length, conformation, the gap between the highest occupied molecular orbital (HOMO) and the lowest unoccupied molecular orbital (LUMO) and the alignment of this gap to the metal Fermi level. Experiments measuring the two terminal conductance of single molecules between metal electrodes have been performed employing a variety of different techniques [2-7]. Although some insight into the transport mechanism has been demonstrated in experiments on devices formed with a large number of molecules in the junction [8, 9], probing the details of the transport mechanism through single molecule devices has proven difficult, primarily due to the large variability in the conductance measurements of such devices [10]. Using amine-gold link chemistry [11], we have previously demonstrated that we can measure the conductance of single molecules reliably and reproducibly and relate the measured conductance to molecular conformations [12]. Here we use chemical substituents to provide new insight to the nature of electron transport through single molecule devices. By studying a series of substituted 1,4 diaminobenzene molecules which have the same length and conformation, we probe the details of the Fermi level alignment to the molecular orbitals.

We measure the conductance of single molecule junctions by breaking Au point contacts in a 1,2,4-trichlorobenzene solution of molecules [6] (See references [10] and [11] for a detailed description.) Conductance traces measured as a function of tip-sample displacement reveal quantized conductance steps observed at multiples of $G_0$ ($2e^2/h$), the fundamental quantum of conductance. In addition, many of the traces reveal steps at a molecule dependent conductance values below $G_0$ (Figure 1A). These steps are due to conduction through a single molecule bridging the gap between the two Au point-contacts. Repeated measurements give a statistical assessment of the junction properties presented as conductance histograms with the peak in the histogram representing the most probable measured conductance value for the molecular junction. Figure 1B shows the conductance histograms for two molecules: 2-chloro 1,4 diaminobenzene and 2-methoxy 1,4-diaminobenzene constructed from thousands of consecutively measured conductance traces without any data selection or processing [11, 12]. These histograms show clear peaks both at multiples of $G_0$, and at a molecule dependent value below $G_0$.

The shift in the peak positions in Figure 1B shows the change in the conductance of the two molecules, differing only in their substituents on the benzene ring. We have performed the same experiment with nine other substituted 1,4 diaminobenzenes (listed in Table 1) and find that their conductance depends on the nature of the substituent. For all the substituents tested, we have also performed control experiments on molecules with the substituent groups at the 1 and 4 positions (for example with 1,4 dicyano benzene) to ensure that the substituent groups does not bind to broken gold point-contacts. For all the substituents listed in Table 1, the histograms measured with the 1,4 disubstituted benzenes do not differ from the control histograms in solvent alone. However for carboxylic group (COOH) and nitro group ($NO_2$), we find that the histograms measured with the 1,4 disubstituted benzenes differ considerably from the control histograms in clean solvent indicating that these groups bind to gold and hence we do not include these in our measurements. For the 1,4 diaminobenzene molecules with substituents, we measure an increase in molecular junction conductance with electron donating substituents (for example with a methoxy or methyl group) as listed in Table 1. For electron withdrawing substituents (for example with a chlorine or a cyano group), we measure a decrease in molecular junction conductance. In order to determine the ultimate physical reason for this increased conductance it is useful to look at the physical changes in the 1,4-diaminobenzene system that accompany the replacement of the ring H atoms with other chemical functional groups.

The highest-occupied molecular orbital in 1,4-diaminobenzene is best described as a combination of the lone-pairs on each of the N atoms and some component of p-$\pi$ density on each of the two C atoms to which the N atoms are bonded. As electron-donating substituents replace H atoms on the ring, the energy of the HOMO orbital increases. When an H is replaced by a methoxy group ($OCH_3$), the $O(2p\pi)$ lone pair delocalizes into the benzene $\pi$ space, thereby raising the energy of the HOMO. Similarly, when electron-withdrawing substituents replace H atoms, the energy of the HOMO is lowered. When H is replaced by Cl, the more electronegative Cl removes electron density from the $\sigma$-space of the benzene, thereby deshielding the $\pi$-space and lowering the energy of the HOMO. Such shifts in the isolated molecule are measured as changes in the ionization potential (IP): electron donating substituents decrease the IP while electron withdrawing substituents increase IP.

To explore this further, we plot in Figure 2A, the measured molecular conductance for the series tested against the vertical IP calculated for the isolated molecules based on density functional theory (DFT) [13-15]. (See supporting information document for details.) There is very limited experimental data available across the series of molecules under study for a comparison. The DFT calculations moderately underestimate the measured IP, as also observed for other aromatics [16] but represent the trends. For example, the measured IP is 6.87 ± 0.05 eV for the unsubstituted 1,4 diaminobenzene and 6.43 eV for tetramethyl 1,4 diaminobenzene [17] while the calculated adiabatic IPs are 6.39 eV and 5.96 eV for the two molecules respectively. We see in Figure 2A that the molecular junction conductance decreases with increasing ionization potential, thus for molecules with a more deeply bound HOMO the conductance is reduced.

We have demonstrated previously[12] that the low bias conductance through polyphenyls attached to gold electrodes amine end-groups is through non-resonant tunneling process or super-exchange. Two different models are frequently used to describe this process when the voltage applied across the junction is much smaller than the HOMO-LUMO gap of the bridging molecule. The molecule can be thought of as a tunnel barrier and the transport details can be understood in terms of Simmons' model [8, 9, 18, 19] as illustrated in Figure 3A. Equivalently, conductance can be related to the electron transfer rate between a donor electrode and an acceptor electrode through a molecular bridge (a model introduced by McConnell [20]), as illustrated in Figure 3B [21, 22]. As long as the metal Fermi level, or equivalently the donor and acceptor levels are far enough from the molecular levels (HOMO and LUMO), the electron tunneling rates decrease exponentially with increasing molecule length ($L$) with the measured low bias conductance $G$ scaling as $e^{-\beta L}$. Here, the decay constant $\beta$ depends on the intrinsic electronic properties of the molecule in the junction. For example, $\beta$ is 0.77/Å for fully saturated diaminoalkanes measured between gold electrodes [11], while it is 0.4/Å for conjugated diamino-polyphenyls [12].

The tunneling process is dominated by the molecular level that is closest to the metal Fermi level, for example, the HOMO, as shown in Figures 3A and B. In this case, the tunneling decay rate $\beta$ depends on the energy separation $\Phi_B=E_{Fermi}-E_{HOMO}$. Although the details of the relation between $\beta$ and $\Phi_B$ differ for the two models ($\beta \propto \Phi_B^{1/2}$ in Simmons' model while $\beta \propto \log(\Phi_B)$ in the McConnell model [20]), both models predict an increase in the measured conductance with a decrease in $\Phi_B$ [23]. Our measurements show that electron donating substituents, which shift the HOMO level up (decreasing IP), increase the junction conductance hence they must decrease $\Phi_B$. Similarly, electron withdrawing substituents must increase $\Phi_B$. These results therefore show that the HOMO is the molecular level that is closest to the Au Fermi level in these molecular junctions. With reference to the Simmons model, transport in these junctions is equivalent to hole tunneling through an insulating film.

This picture is further supported by quantum chemistry calculation [13-15] for all molecules studied with the amine link groups coupled to Au clusters representing the contacts [11, 12]. The binding of an Au atom to each amine results in frontier orbitals that are predominantly of Au-s, N lone pair anti-bonding character. These frontier orbitals are tunnel coupled through the molecular backbone resulting in a symmetric and anti-symmetric pair with a splitting, *2t*. In the cases with a single substituent, the molecule has a built-in dipole which breaks the symmetry between the two Au contact atoms. However, the dipole is oriented nearly orthogonal to the junction (the Au-Au contact line) and the degree of asymmetry is small, as judged from the frontier molecular orbitals.[24] In figure 2B, we plot $4t^2$, which is a proportional to the molecule conductance [11, 12, 20, 25], against the ionization potential (also listed in Table 1). A comparison of Figures 2A and B indicate that the trends in the measured conductance are accounted for by the changes in the tunnel coupling of the frontier orbitals across the molecular backbone induced by the substituents. Furthermore, we find a close correspondence in the iso-surface plot of the frontier orbital (Figure 3D) and the HOMO level for the isolated molecule (Figure 3C). This provides additional evidence that it is indeed the HOMO that is the molecular orbital that is closest to the Fermi level.

We have shown that the conductance of diaminobenzene is altered by having substituents on the benzene ring. It is well known that reaction rates can be affected by substituents. These effects are often quantified by use of substituent constants, an idea that was first introduced by L. P. Hammett.[26] Reaction rates for a substituted molecule relative to the rate for an un-substituted molecule are plotted against the Hammett substituent constant in order to gain insight into the reaction mechanism. In organic reactions, the logarithm of the reaction rate is proportional to the activation energy for the reaction and the Hammett equation is an example of a linear free energy relation. Extending this idea to electron tunneling, we can relate the molecular conductance to the Hammett constants as the logarithm of conductance is a function of the tunneling barrier height.

In Figure 4, we plot the log of the ratio of the measured conductance for the substituted ($G_X$) and un-substituted ($G_H$) molecule scaled by the number of substituents on the ring against the Hammett parameter $\sigma_{para}$ from reference [27]. We have made three major assumptions here. First, the choice of Hammett constant $\sigma_{para}$ may not represent the effect of the substituent fully. Second, we assume that the effect of multiple substituents is additive, even though it is clear from the methyl and fluorine substituted data that this is not so. Third, we know that the logarithm of the conductance is not linear in the tunnel barrier, although for small conductance changes measured here, the deviation from a linear relation is within our error bar. Nonetheless, Figure 4 shows a clear trend, that the conductance decreases as the Hammett constant is made more positive. This negative slope indicates that the transition state in this tunneling "reaction" is positively charged. Quite remarkably, this is consistent with the picture of hole tunneling described above.

In conclusion, we have measured the conductance of diamino benzene with different substituents that shift the molecular levels relative to the Au Fermi level in the single molecule junctions measured. We find that electron donating substituents increase conductance while electron withdrawing substituents decrease conductance resulting in a chemical gating of the junction conductance. These results provide evidence that the molecular HOMO level is closer to the Au Fermi.

**Acknowledgments:** We thank Jen Klare, Horst Stormer, and Jim Yardley for useful discussions. This work was supported primarily by the Nanoscale Science and Engineering Initiative of the National Science Foundation under NSF Award Number CHE-0117752 and by the New York State Office of Science, Technology, and Academic Research (NYSTAR). M.L.S. thanks the Material Research Science and Engineering Center program of the NSF under award number DMR-0213574.

Table 1: List of the molecules studied showing the number and type of substituents, measured conductance histogram peak position, calculated IP and calculated relative conductance. The calculated conductance is determined by scaling the calculated tunneling probability by the measured peak conductance of the 1,4 diaminobenzene.

| Number | Molecule Name | Substituent | Calculated IP (eV) | Conductance Peak ($\times 10^{-3}\,G_0$) | Calculated Relative Conductance ($\times 10^{-3}\,G_0$) |
|---|---|---|---|---|---|
| 1 | Tetramethyl 1,4 diaminobenzene | $CH_3$ ($\times$ 4) | 6.36 | 8.2 ± 0.2 | 7.6 |
| 2 | 2,5 Dimethyl 1,4 diaminobenzene | $CH_3$ ($\times$ 2) | 6.59 | 6.9 ± 0.2 | 6.7 |
| 3 | 2-Methoxy 1,4 diaminobenzene | $OCH_3$ ($\times$ 1) | 6.55 | 6.9 ± 0.2 | 7.1 |
| 4 | 2-Methyl 1,4 diaminobenzene | $CH_3$ ($\times$ 1) | 6.72 | 6.4 ± 0.6 | 6.5 |
| 5 | 1,4 diaminobenzene | H ($\times$ 4) | 6.83 | 6.4 ± 0.2 | 6.4 |
| 6 | 2,5 Dichloro 1,4 diaminobenzene | Cl ($\times$ 2) | 7.14 | 6.1 ± 0.2 | 6.0 |
| 7 | 2-Bromo 1,4 diaminobenzene | Br ($\times$ 1) | 7.02 | 6.1 ± 0.6 | 6.1 |
| 8 | Trifluoromethyl 1,4 diaminobenzene | $CF_3$ ($\times$ 1) | 7.22 | 6.1 ± 0.2 | 6.2 |
| 9 | 2-Chloro 1,4 diaminobenzene | Cl ($\times$ 1) | 7.00 | 6.0 ± 0.4 | 6.2 |
| 10 | 2-Cyano 1,4 diaminobenzene | CN ($\times$ 1) | 7.30 | 6.0 ± 0.3 | 5.9 |
| 11 | 2-fluoro 1,4 diaminobenzene | F ($\times$ 1) | 7.03 | 5.8 ± 0.4 | 6.3 |
| 12 | Tetrafluoro 1,4 diaminobenzene | F ($\times$ 4) | 7.56 | 5.5 ± 0.3 | 5.2 |

**Figure Captions:**

Figure 1: (A) Sample conductance traces measured in the presence of molecule **3** shown on a semi-log scale. The traces are arbitrarily offset along the x-axis (relative tip-sample displacement). (B) Conductance histogram of **3** (blue) and **9** (red) on a log scale. Peaks are clearly visible at $G_0$ and $2 \times G_0$ for both curves and at a substituent dependent value below $G_0$. Inset (lower): Illustration of a gap between a gold point-contact that can be bridged with a substituted benzene diamine. Inset (upper): Lorentzian fits to the molecular conductance peaks for both histograms shown on a linear scale. The bin size is $10^{-4}\ G_0$ for all histograms and data were taken at 25 mV.

Figure 2: (A) Measured conductance values against the calculated ionization potential for the series of 11 molecules tested. The number of traces measured ranged from 12000 to 30000 for different molecules. For each molecule, histograms of 1000 consecutive traces were computed and a Lorentzian was fit to the molecular peak. The mean and standard deviation of the peak positions determined the molecule conductance and the error bar (also listed in Table 1). (B) Square of the calculated tunnel coupling ($4 \times t^2$) against the calculated ionization potential.

Figure 3: (A) Illustration of the Simmon's model used to describe a metal-molecule-metal junction, where the HOMO level is shown closer to the metal Fermi level. L is the molecule length, and $\Phi_B$ is the energy difference between the molecular HOMO level and the metal Fermi level. (B) Illustration of the McConnell model used to describe electron transfer through a metal donor – molecular bridge- metal acceptor junction. $V_{DB}$ and $V_{BA}$

are the coupling between the donor level or acceptor level and the closest molecular orbital, $V_{BB}$ is the coupling between adjacent the molecular orbitals. For the specific case of a diaminobenzene linked to two gold atoms, $V_{DB}$ and $V_{BA}$ refer to the Au-N lone pair coupling, and $V_{BB}$ represents C-C or C-N coupling. (C) Iso-surface plot of the HOMO level for an isolated diaminobenzene molecule and for Au-s orbitals which couple to the N-lone pair. (D) Iso-surface plot of the frontier orbital responsible for tunneling obtained from DFT calculations of diaminobenzene coupled to two gold atoms.

Figure 4: Logarithm of the ratio of the substituted molecule conductance ($G_X$) to the unsubstituted molecule conductance ($G_H$) divided by the number of substituents (N) in each molecule against the Hammett $\sigma_{para}$ coefficient from reference.[27]

Figure 1

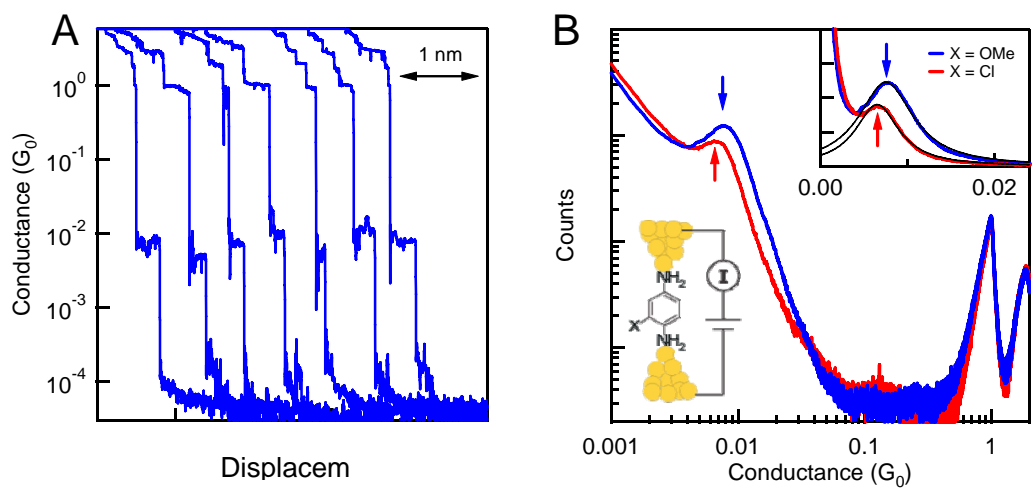

Figure 2

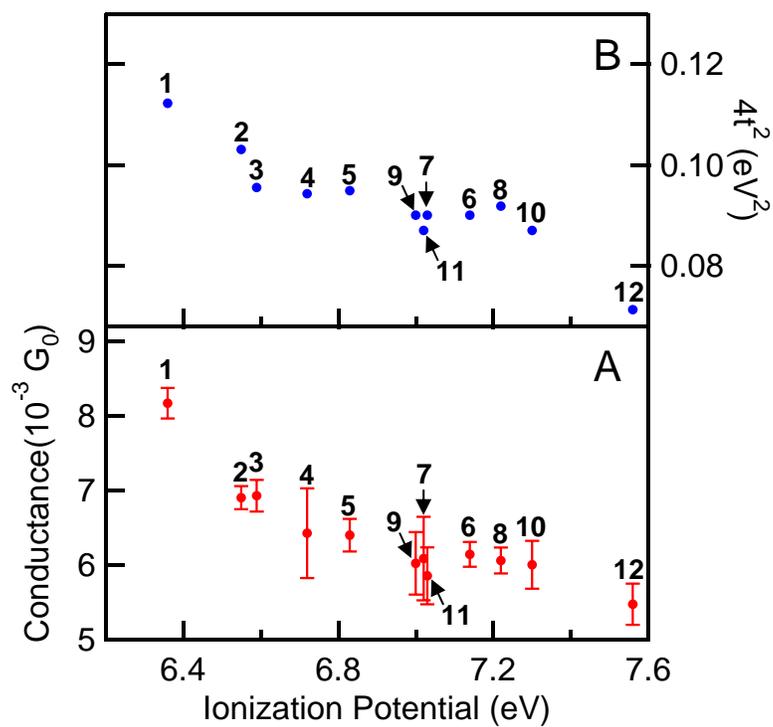

Figure 3

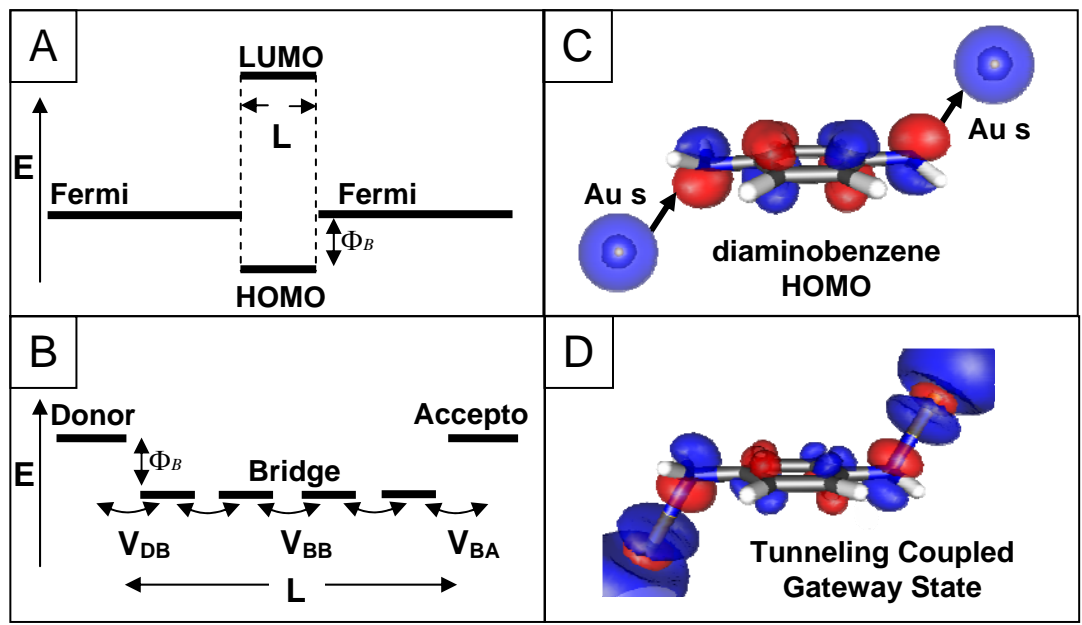

Figure 4

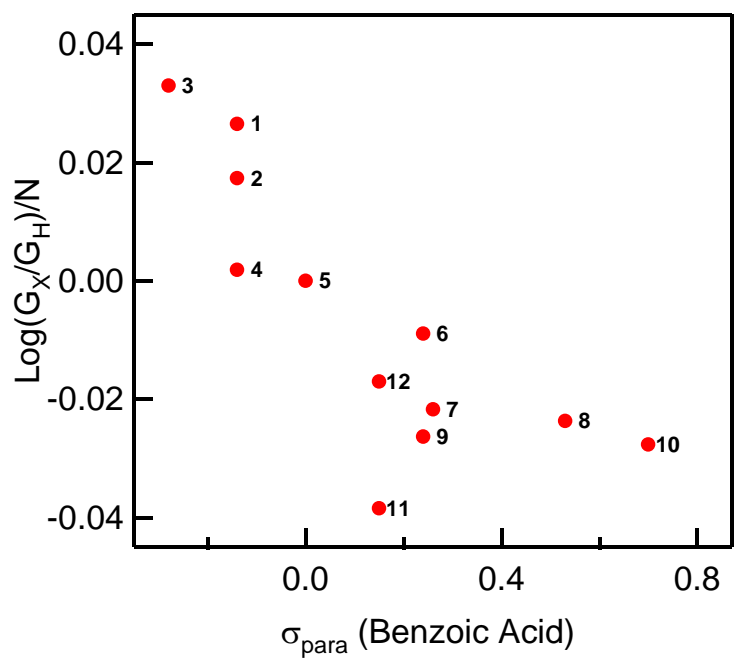